\documentclass[aps,prb,twocolumn,superscriptaddress,showpacs,floatfix]{revtex4-1}
\usepackage{amsmath,amssymb,graphicx}

\usepackage[utf8]{inputenc}
\usepackage[T1]{fontenc}
\usepackage{xcolor}

\IfFileExists{newtxtext.sty}
   {\usepackage{newtxtext,newtxmath}}
   {\IfFileExists{stix.sty}
      {\usepackage{stix}}
      {\IfFileExists{mathptmx.sty}
      {\usepackage{mathptmx}}{} } }

\usepackage{textcomp}

\usepackage{bm}

\IfFileExists{siunitx.sty}{\usepackage{booktabs,siunitx}}{}

\pdfoutput=1
\usepackage{color}
\definecolor{LinkColor}{rgb}{0.256,0.439,0.588}
\usepackage{hyperref}
\hypersetup{
   pdfauthor={Zi-Hong Liu, Xiao Yan Xu, Yang Qi, Kai Sun, and Zi Yang Meng},
   pdftitle={Itinerant quantum critical point with frustration and non-Fermi-liquid},
   colorlinks=true,
   citecolor=LinkColor,
   linkcolor=LinkColor,
   urlcolor=LinkColor
}

\renewcommand{\vec}[1]{\mathbf{#1}}

\usepackage{pifont}

\newcolumntype{L}[1]{>{\raggedright\let\newline\\\arraybackslash\hspace{0pt}}m{#1}}
\newcolumntype{C}[1]{>{\centering\let\newline\\\arraybackslash\hspace{0pt}}m{#1}}
\newcolumntype{R}[1]{>{\raggedleft\let\newline\\\arraybackslash\hspace{0pt}}m{#1}}

\begin{document}

\title{Itinerant quantum critical point with frustration and non-Fermi-liquid}

\author{Zi Hong Liu}
\affiliation{Beijing National Laboratory for Condensed Matter Physics and Institute
of Physics, Chinese Academy of Sciences, Beijing 100190, China}
\affiliation{School of Physical Sciences, University of Chinese Academy of Sciences, Beijing 100190, China}
\author{Xiao Yan Xu}
\affiliation{Beijing National Laboratory for Condensed Matter Physics and Institute of Physics, Chinese Academy of Sciences, Beijing 100190, China}
\affiliation{School of Physical Sciences, University of Chinese Academy of Sciences, Beijing 100190, China}
\affiliation{Department of Physics, Hong Kong University of Science and Technology, Clear Water Bay, Hong Kong, China}
\author{Yang Qi}
\affiliation{Department of Physics, Fudan University, Shanghai 200433, China}
\affiliation{Department of physics, Massachusetts Institute of Technology, Cambridge, MA 02139, USA}
\author{Kai Sun}
\affiliation{Department of Physics, University of Michigan, Ann Arbor, MI 48109, USA}
\author{Zi Yang Meng}
\affiliation{Beijing National Laboratory for Condensed Matter Physics and Institute of Physics, Chinese Academy of Sciences, Beijing 100190, China}
\affiliation{CAS Center of Excellence in Topological Quantum Computation and School of Physical Sciences,
University of Chinese Academy of Sciences, Beijing 100190, China}

\begin{abstract}
Employing the self-learning quantum Monte Carlo algorithm, we investigate the frustrated transverse-field triangle-lattice Ising model coupled to a Fermi surface. Without fermions, the spin degrees of freedom undergoes a second-order quantum phase transition between paramagnetic and clock-ordered phases. This quantum critical point (QCP) has an emergent U(1) symmetry and thus belongs to the (2+1)D XY universality class. In the presence of fermions, spin fluctuations introduce effective interactions among fermions and distort the bare Fermi surface towards an interacting one with hot spots and Fermi pockets. Near the QCP, non-Fermi-liquid behavior are observed at the hot spots, and the QCP is rendered into a different universality with Hertz-Millis type exponents. The detailed properties of this QCP and possibly related experimental systems are also discussed.
\end{abstract}

\date{\today}

\maketitle

\section{Introduction}
Quantum criticality in correlated itinerant electron systems is a subject with great theoretical and experimental significance~\cite{Hertz1976,Millis1993,Moriya1985,Stewart2001,Chubukov2004,Belitz2005,Loehneysen2007,Chubukov2009}, and plays a vital role in the study of anomalous transport, strange metal and non-Fermi-liquid~\cite{Metzner2003,Senthil2008,Holder2015,Metlitski2015} in heavy-fermion materials~\cite{Custers2003,Steppke2013}, cupurates and Fe-based high-temperature superconductors~\cite{ZhangWenLiang2016,LiuZhaoYu2016,Gu2017}. Among its many interesting aspects, the fate of the Fermi surface (FS) and the nature of low-energy excitations in the quantum critical region are of particular importance. 

In this paper, we focus on one family of itinerant QCPs, where order parameters have finite wave vectors, such as anti-ferromagnetism (AFM) and charge- or spin- density wave (CDW/SDW) states. Although extensive efforts have been devoted, theoretical understanding about these QCPs has not yet reached convergence, due to its nonperturbative nature. For example, according to the Hertz-Millis-Moriya theory~\cite{Hertz1976,Millis1993,Moriya1985}, such a QCP in 2D is characterized by mean-field scaling exponents with dynamic critical exponent $z=2$, and renormalization group (RG) analysis reveals linear temperature ($T$) dependence in spin susceptibility with logarithmic corrections. However, as higher order contributions are taken into account, theories beyond Hertz-Millis-Moriya have been proposed and extensively studied~\cite{Abanov2003,Abanov2004,Metlitski2010,Sur2016,Schlief2017,SSLee2018,Schlief2018}, where novel phenomena, e.g. anomalous dimensions, are expected to emerge.
  Recently, the development in sign-problem-free quantum-Monte-Carlo (QMC) methods provides a new path way to sharpen our understanding about these open questions, and this unbiased numerical technique has been utilized to study various itinerant QCPs including Ising-nematic~\cite{Schattner2015a,Lederer2016}, ferromagnetic~\cite{Xu2016}, CDW~\cite{ZXLi2015} and SDW~\cite{Berg12,ZXLi2016,Schattner2015b,Gerlach2017}. For SDW QCPS, the signature of a $z=2$ has been observed~\cite{Berg12,ZXLi2016,Schattner2015b,Gerlach2017}, but the linear $T$ dependence predicted in the Hertz-Millis-Moriya theory and the possible anomalous dimensions have not yet been explored in QMC studies. In contrast to the conclusion of $z=2$, which comes from the leading-order random phase approximation (RPA),  these two predictions rely on higher order effects and RG flows, and thus will only arise in the close vicinity of the QCP, which are challenging for QMC simulations to explore, because (1) in most previous studies, the QCPs turn out to be covered by a superconducting dome, which prevents the access to the quantum critical region and (2) the diverging correlation length near a QCP enhances the finite-size effect, thus requiring larger system sizes to obtain reliable results.

To overcome these difficulties, we utilize two new methodologies in this paper. (1) We designed a new model with two identical copies of fermions. As have been demonstrated previously, such a double-copy construction greatly suppress superconductivity near the ferromagnetic QCP~\cite{Xu2016}. Here, using the same general idea, we find a pristine itinerant SDW QCP without superconductivity or any other competing orderings to interrupt the critical scaling. (2) By employing a self-learning quantum Monte Carlo algorithm~\cite{Xu2016self,liu2016self,liu2016fermion,Nagai2017,HTShen2018,ZHLiu2018,Chen2018} to improve numerical efficiency, our simulations can approach larger system sizes, and thus the finite-size effect is under control. In the close vicinity of the QCP, our numerical data supports the Hertz-Millis-Moriya prediction, including the linear $T$ dependence~\cite{Millis1993}, and gives a very small upper bound on the numerical value of anomalous dimension.

\begin{figure*}[t!]
\includegraphics[width=\textwidth]{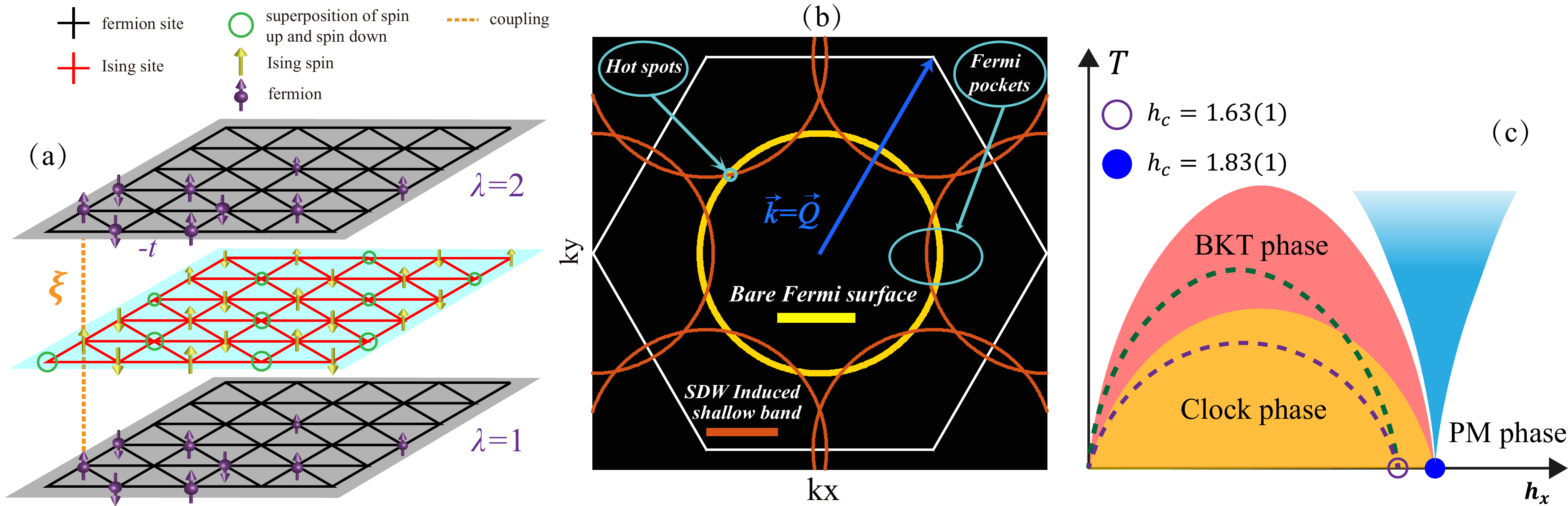}
\caption{(a) Illustration of our model. Fermions reside on two of the layers ($\lambda$ = 1,2) with intra-layer nearest-neighbor hopping $t$. The middle layer is composed of Ising spins $s^{z}_{i}$, subject to nearest-neighbor antiferromagnetic Ising coupling $J$ and a transverse magnetic field $h$. Between the layers, an onsite Ising coupling is introduced between fermion and Ising spins ($\xi$). 
(b) The bare FS of the $H_{\text{fermion}}$ (yellow circle) and the folded FS (orange circles), coming from translating the bare FS by momentum $\mathbf{Q}$ (blue arrow). The folded FS contains Fermi pockets and hot spots. (c) Semi-quantitative phase diagram. The dashed lines mark the phase boundaries of the naked bosonic model $H_{\text{spin}}$, with a QCP (open magenta dot) at $h_c=1.63(1)$~\cite{Isakov2003,Wang2016}. The filled areas are the phases with fermions. The orange area is the clock phase with long-range order of Ising spins and Fermi pockets and hot spots, the pink area is the BKT phase with power-law correlation functions, the blue area is the quantum critical region , and the white area is the disordered paramagnetic phase. The QCP (solid blue dot) is shifted to a higher value $h=1.83(1)$ in comparison to the naked bosonic one, where non-fermi-liquid behavior emerges near the hot spots.}
\label{fig:model}
\end{figure*}

It is also worthwhile to highlight that in comparison with previous studies~\cite{Schattner2015a,Lederer2016,ZXLi2015,Berg12,ZXLi2016,Schattner2015b,Gerlach2017}, our model,  which resides on a triangular lattice,  is bestowed with the new ingredient of geometric frustration. The frustration results in an emergent U(1) symmetry, as well as a Berezinskii-Kosterlitz-Thouless (BKT) phase. On the experimental side, our findings have immediate relevance towards the understanding of the recent experiments in frustrated itinerant systems, such as anomalous transport in rare-earth triangular lattice antiferromagnet CeCd$_3$As$_3$~\cite{YQLiu2016} and frustrated Ising-like heavy-fermion compound CePdAl~\cite{HengcanZhao2016}. And the  newly discovered transition-metal superconductor families, CrAs~\cite{Wu2014}, MnP~\cite{Cheng2015} and CrAs$_{1-x}$P$_x$~\cite{Cheng2017}, where non-Fermi-liquid behavior close to the itinerant antiferromagnetic quantum critical point has been observed.

\section{Model and Method} 
Our model is defined on layered triangular lattice with
\begin{equation}
H = H_{\text{fermion}} + H_{\text{spin}} + H_{\text{f-s}}.
\end{equation}
As shown in Fig.~\ref{fig:model} (a), fermions, described by
\begin{equation}
H_{\text{fermion}}=-t\sum_{\langle i,j \rangle,\lambda,\sigma}(c^{\dagger}_{i,\lambda,\sigma}c_{j,\lambda,\sigma}+\text{h.c.})-\mu\sum_{i}n_{i},
\end{equation}
subject to intra-layer nearest-neighbor hopping $t$ and chemical potential $\mu$, reside on two of the layers $\lambda=1,2$. The middle layer, decribed by 
\begin{equation}
H_{\text{spin}}=J\sum_{\langle i,j \rangle}s^{z}_{i}s^{z}_{j}-h\sum_{i}s^{x}_{i},
\end{equation}  
is composed of Ising spins $s^{z}_{i}$ with frustrated antiferromagnetic Ising coupling $J>0$ and a transverse magnetic field $h$ along $x$. Fermions and Ising spins are coupled together via an inter-layer onsite Ising coupling 
\begin{equation}
H_{\text{f-s}}=-\xi\sum_{i}s^{z}_{i}(\sigma^{z}_{i,1}+\sigma^{z}_{i,2}),
\end{equation} 
where $\sigma^{z}_{i,\lambda}=\frac{1}{2}(c^{\dagger}_{i,\lambda,\uparrow}c_{i,\lambda,\uparrow}-c^{\dagger}_{i,\lambda,\downarrow}c_{i,\lambda,\downarrow})$ is the fermion spin along $z$ and $\xi$ is the coupling strength. We set $t=1$, $J=1$, $\mu=-0.5$ (electron density $\langle n_{i,\lambda}\rangle \sim 0.8$) and leave $h$ and $\xi$ as control parameters.

$H_{\text{spin}}$ describes a frustrated triangular-lattice transverse-field Ising model with extensive ground state degeneracy at $h=0$. At finite $h$, this degeneracy is lifted by the quantum order-by-disorder effect, resulting in an ordered ground state with clock pattern~\cite{Moessner2001}. As shown in the middle layer of Fig.~\ref{fig:model} (a), the clock phase  breaks spontaneously the translational symmetry~\cite{Isakov2003,Wang2016} and thus has an enlarged unit cell with three sublattices. This phase is characterized by a complex order parameter $me^{i\theta}=m_{1}+m_{2}e^{i4\pi/3}+m_{3}e^{-i4\pi/3}$ where $m_{\alpha}=\frac{1}{3N}\sum^{N/3}_{i=1} s^{z}_{i,\alpha}$ with $\alpha=1,2,3$ representing magnetization of the three sublattices. 
In the momentum space, this order parameter has a finite wave-vector $\mathbf{k}=\mathbf{Q}=(\frac{2\pi}{3},\frac{2\pi}{\sqrt{3}})$ 
as shown in Fig.~\ref{fig:model}(b). Upon introducing quantum/classical fluctuations via increasing $h$ or $T$, the ordered phase can melt. The quantum melting is through a second-order quantum phase transition at $h_c=1.63(1)$ with an emergent U(1) symmetry~\cite{Wang2016}. 
Because of this emergent continuous symmetry, despite that $H_{\text{spin}}$ describes an Ising model, this quantum critical point belongs to the (2+1)D XY universality class and the thermal melting of the clock phase involves an intermediate BKT phase~\cite{Wang2016}.

In the presence of the fermion-spin coupling, which is relevant in RG sense, the three phases survive with shifted phase boundaries as shown in Fig.~\ref{fig:model}(c). Furthermore, because fermion and Ising spins are coupled together, the Ising-spin clock phase immediately generates a SDW ordering in the fermionic sector with finite ordering wavevector $\mathbf{Q}$, which folds the Brillouin zone and renders a new Fermi surface (FS) with pockets as schematically shown in Fig.~\ref{fig:model}(b). Near the QCP, the quasi-particle at the tip of the FS pockets lose their coherence, forming the so-called hot spots. 

\begin{figure}[htp]
\centering
\includegraphics[width=\columnwidth]{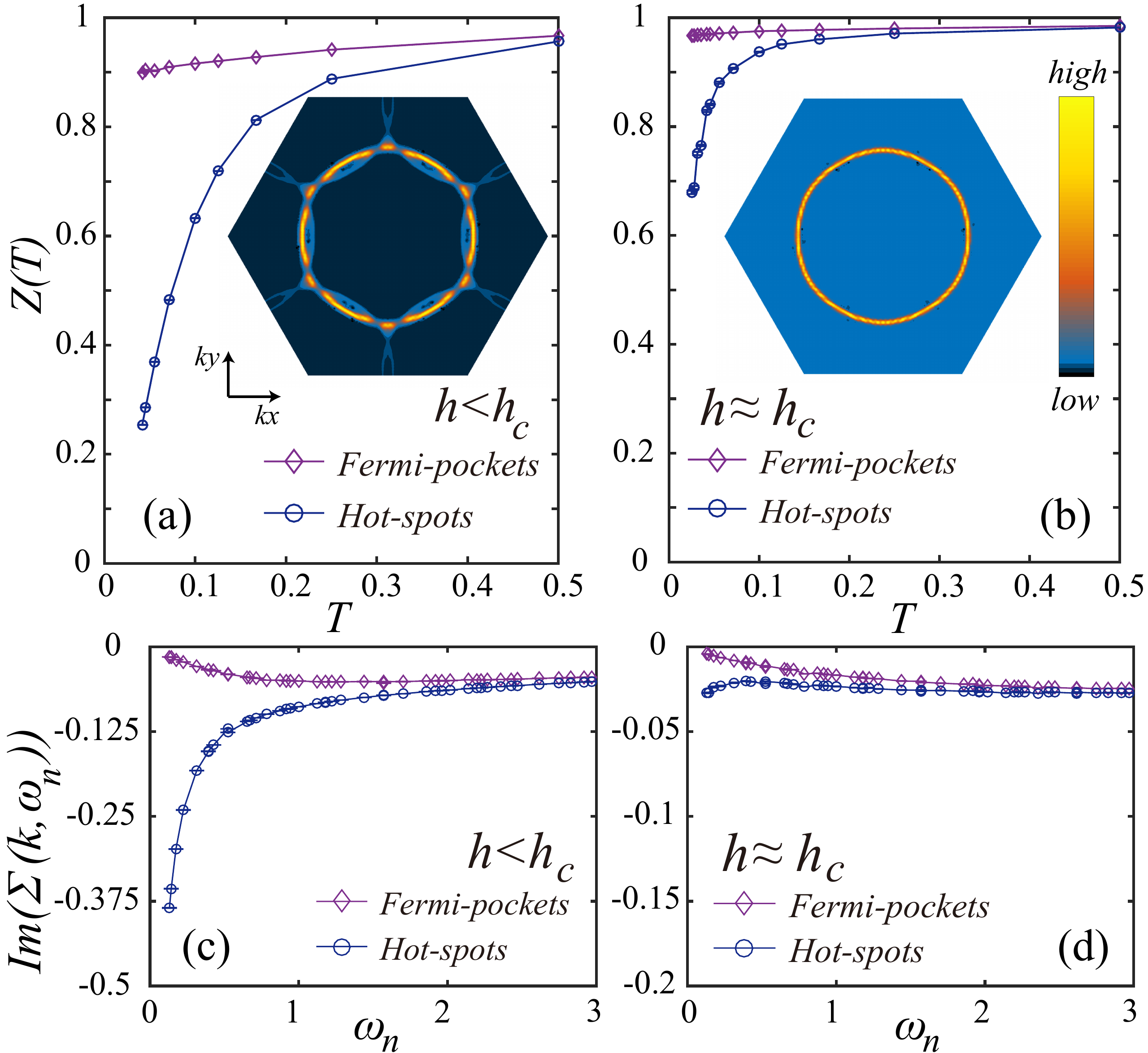}
\caption{Quasiparticle weight $Z(T)$ [(a) and (b)] and fermion self-energy $\Sigma(\mathbf{k},\omega_n)$ [(c) and (d)]. (a) and (c) shows the ordered phase $h=1.0<h_c$, while (c) and (d) are at the QCP $h\approx h_c$. In (a) and (c), $Z(T)$ is suppressed on the hot spot due to the gap opening and correspondingly $\text{Im}\Sigma(\mathbf{k},\omega_n)$ diverges, whereas in (b) and (d), $Z(T)$ is suppressed at the hot spot due to the emergence of non-Fermi-liquid behavior and correspondingly the $\text{Im}\Sigma(\mathbf{k},\omega_n)$ saturate at low frequency.}
\label{fig:FS}
\end{figure}

To unveil this process with unbiased numerical approach, we performed self-learning determinantal quantum Monte Carlo simulations~\cite{Xu2016self,Chen2018}. While  details are given in Appendix~\ref{sec:SLDQMCIntroduction}, here we highlight two key aspects of the method: i) to update the configuration weight -- comprised of the contribution from both Ising spin Boltzmann factor and fermion determinant --  
more effectively, we implement cluster update~\cite{Xu2016a} as well as the cumulative updates in self-learning Monte Carlo~\cite{liu2016self,liu2016fermion,Xu2016self,Nagai2017,ZHLiu2018,Chen2018}. This advanced Monte Carlo technique enable us to comfortably simulate systems as large as $L=30$ and temperature as low as $\beta t=40$, and overcome the critical slowing down in the vicinity of QCP. ii)  the two layer construction of fermions guarantees that our Hamiltonian is invariant under an anti-unitary (time-reversal followed by an orbital rotation $i\tau_{y}$) transformation, so that we can simulate this system without sign problem at any filling~\cite{Xu2016}.

\section{Locating QCP and non-fermi liquid at hot-spots} 
To locate the QCP, we measure both correlation ratio~\cite{Pujari2016} and binder ratio of the Ising spins as a function of $h$. As shown in Fig.~\ref{fig:Rc-ratio} in  Appendix~\ref{sec:correlationratio} and Fig.~\ref{fig:superfluid-density} in Appendix~\ref{sec:superfluiddensity}, the crossing points of both ratios give rise to $h_c = 1.83(1)$ and no superconductivity emerges near the QCP. In the insets of Figs.~\ref{fig:FS}(a)  and (b), Fermi surfaces at $h=1.0<h_c$ and $h\approx h_c$ are presented respectively. Here, we plot the fermion Green's function $G(\mathbf{k},\frac{\beta}{2})$. At low temperature, $G(\mathbf{k},\frac{\beta}{2})\approx \beta A(\mathbf{k},\omega=0)$~\cite{Schattner2015a,Xu2016} and thus this quantity reveals the spectral function, which peaks at the Fermi surface. At $h<h_c$, the zone folding induced by the Ising-spin clock phase is clearly manifested, with Fermi pockets and hot spots similar as those shown in the schematic plot (Fig.~\ref{fig:model}(b)). At $h\approx h_c$, the zone folding disappears, indicating that the lattice translational symmetry is recovered, but non-Fermi liquid behaviors arise at hot spots. To demonstrate the non-Fermi liquid behavior, we plot in Figs.~\ref{fig:FS}(a) and (b), the temperature dependence of quasiparticle weight calculated as $Z_{\mathbf{k}_{F}}(T) = \frac{1}{1-\frac{\text{Im}\Sigma(\mathbf{k}_{F},i\omega_{0})}{\omega_{0}}}$, where $\omega_0=\pi T$ is the lowest Matsubara frequency~\cite{Chen2012}, and in Figs.~\ref{fig:FS}(c) and (d) the corresponding fermion self-energy $\text{Im}\Sigma(\mathbf{k},\omega)$, with $\mathbf{k}$ at the Fermi pocket and at the hot spot. At $h<h_c$ (Fig.~\ref{fig:FS} (a)), the quasiparticles on the Fermi pocket are well-defined with $Z(T)$ close to unity and vanishing self-energy at low $T$. At the hot spots, due to the gap opening from the zone folding, $Z(T)$ vanishes with divergent Fermi self-energy. Near the QCP ($h\approx h_c$), Fermi liquid behavior is observed near the Fermi pocket with finite $Z(T)$ and vanishing $\Sigma$ at low $T$. At the hot-spots, however, $Z(T)$ is strongly suppressed as $T$ reduces with a finite fermi self-energy, which are key signatures of a non-Fermi liquid.

\begin{figure}[htp]
\centering
\includegraphics[width=\columnwidth]{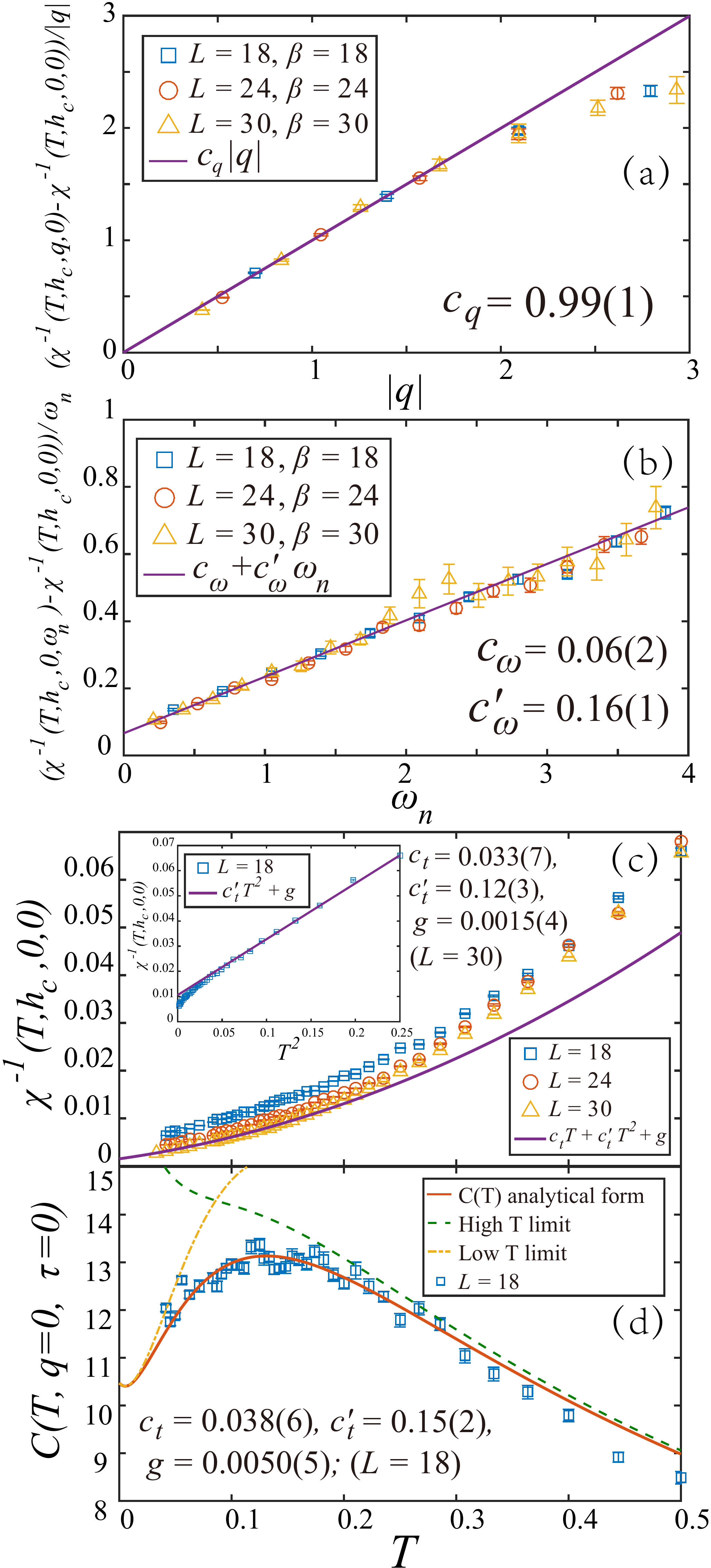}
\caption{Spin susceptibility $\chi(T,h_c,\mathbf{q},\omega)$. To determine the $q$ and $\omega$ dependence, we plot $\frac{\chi^{-1}(\mathbf{q})-\chi^{-1}(0)}{|\mathbf{q}|}$ and $\frac{\chi^{-1}(\omega)-\chi^{-1}(0)}{\omega}$ in (a) and (b). Although linear behaviors are observed in both figures, the difference in intersections [(a) zero and (b) finite] is crucial and indicating a clean quadratic momentum dependence $c_q|\mathbf{q}|^2$ and a crossover $c_\omega \omega+c_\omega' \omega^2$ frequency dependence. (c) presents the $T$ dependence. At high $T$, the linear relation in the inset ($\chi^{-1}$ vs. $T^2$)  indicates a $T^2$ dependence. However, the low $T$ part  deviates strongly from $T^2$ and fits well to linear $T$. The constant $g$ here is the finite size gap, which scales to zero in the thermodynamic limit. (d) The equal-time correlation function. Here we compared numerical results (dots) and the analytic theory in Eq.~\ref{eq:equal-time}] (red solid line). The theoretical curve contains no adjustable parameters, where all values are determined from Fig.(b) and the $L=18$ results in (c). Dashed lines (yellow and green) show the asymptotic behaviors at low and high $T$ respectively.}
\label{fig:chifit}
\end{figure}

\section{Quantum scaling analysis}
Next we discuss the scaling behavior at the QCP obtained from Ising spin susceptibility, $\chi(T,h,\mathbf{q},\omega_n)$. We define $\mathbf{q}=\mathbf{k}-\mathbf{Q}$ as the relative momentum vector with respect to the position of the hot spots in BZ. 

In the absence of fermions, the QCP is well known~\cite{Isakov2003,Wang2016} in the (2+1)D XY universality class,
and the  Ising spin susceptibility takes the following form~\footnote{For simplicity, the extremely small anomalous dimension ($\eta=0.04$) of (2+1)D XY universality class is ignored here in our theoretical description. Our QMC results also confirm this (details are showed in Appendix~\ref{sec:bosonicsuscepU0}).} near the QCP,
\begin{equation}
\chi= \frac{1}{c_{t}T^{2}+c_{h}|h-h_c|^{\gamma}+c_q |\mathbf{q}|^2 + c_{\omega} \omega^{2}}.
\end{equation} 
This behavior of bare bosonic spin susceptibility is also verified by turning off coupling to fermions in our simulation, as showed in Fig.~\ref{fig:chi-analysis-U0} in Appendix~\ref{sec:bosonicsuscepU0}. 

In the presence of itinerant fermions, the fermion-boson coupling will modify the scaling behavior at the phase transition. Because of its strong coupling nature, to understand this itinerant QCP is a challenging task. In the past decades, various scenarios and approximations have been utilized, which can be largely classified into three categories: (1) to the leading order,  i.e. within the random phase approximation (RPA), the fermion Landau damping introduces an extra linear $\omega$ term to $\chi^{-1}$, which dominates the lower-energy quantum dynamics, transforming the dynamic critical exponent $z$ from $1$ to $2$~\cite{Hertz1976}. (2) Starting from the RPA effective theory, the RG analysis by Millis suggests that the T dependence in $\chi^{-1}$ flows from $T^2$ to $T$ (with logarithmic corrections) at low temperature, $\ln[\ln(1/T)] \gg 1$~\cite{Millis1993}. These results (linear $\omega$ and $T$ dependence) are known as the Hertz-Millis-Moriya theory~\cite{Moriya1985}. Because the effective dimension here is $z+d=4$, the upper critical dimension for a $\phi^4$-theory, anomalous dimensions are not expected within these approximations. However, (3) by further taking into account higher-order fermionic contributions beyond the RPA, additional non-local interactions arises, which has been argued to result in the breakdown of the Hertz-Millis-Moriya theory, e.g., finite anomalous exponent~\cite{Abanov2003,Abanov2004,Metlitski2010,Sur2016}. In this work, our objective is to compare the unbiased numerical results with predictions from all these three categories. Our results show good agreement with the first two, consistent with the Hertz-Millis-Moriya theory, while gives a very small upper bound on the numerical value of the anomalous exponents. The details of our results are presented in the following.

\begin{figure}[htp!]
\centering
\includegraphics[width=\columnwidth]{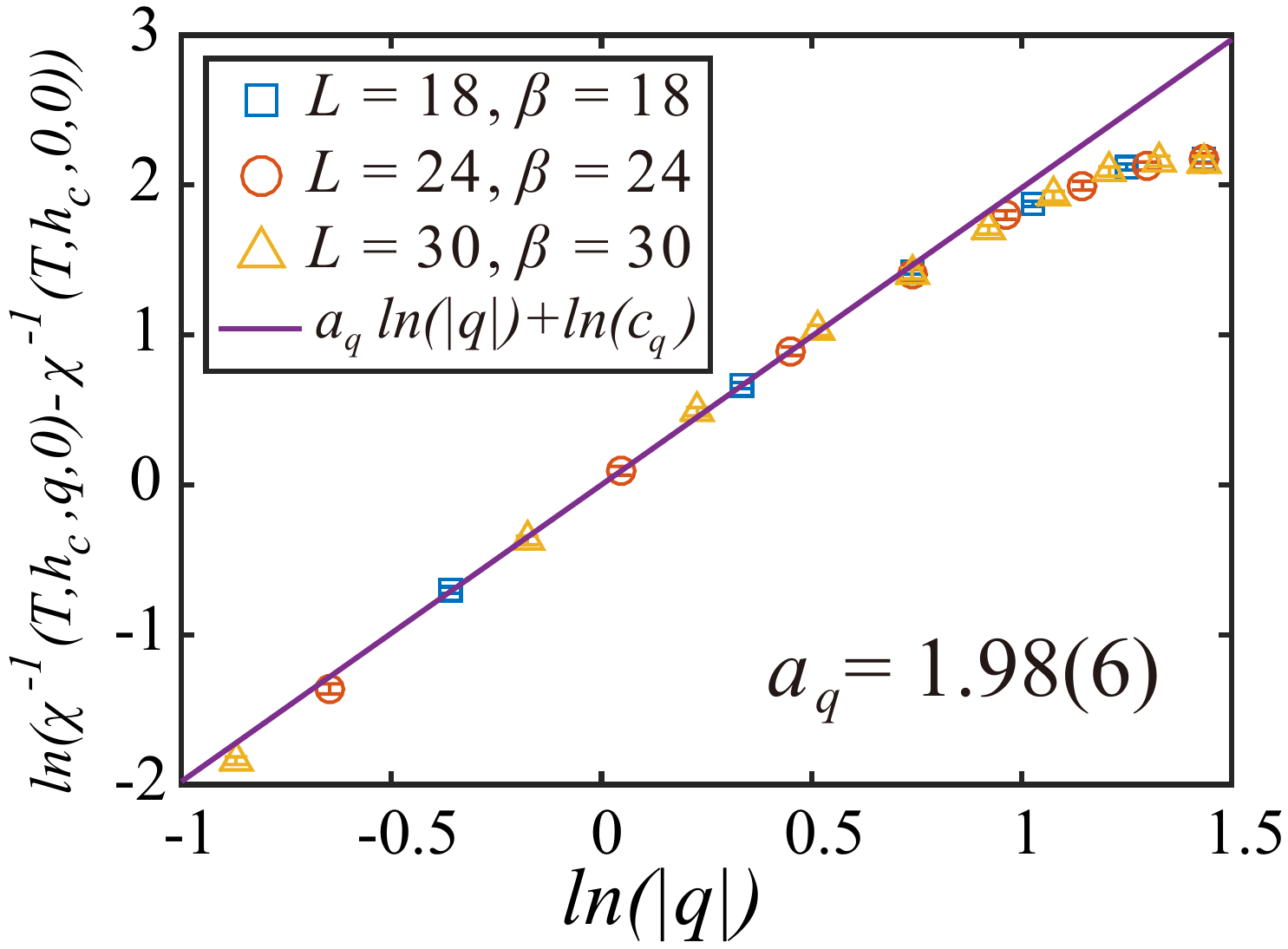}
\caption{We set $h=h_c$, $\omega_n=0$ and plot $\ln(\chi^{-1}(h_c,T,|\mathbf{q}|,0)-\chi^{-1}(h_c,T,0,0))$ as function of $\ln(|\mathbf{q}|)$ to obtain the power-law behavior $c_q|\mathbf{q}|^{a_q}|$ in the momentum dependence of bosonic susceptibility.}
\label{fig:chiqpower}
\end{figure}

In the presence of itinerant fermions, the quantum phase transition is found to remain second order and the spin susceptibility form is found to get corrections from itinerant fermions. 

A controllable quantum scaling analysis is first done by exploring $\vec{q}$, $\omega$, $T$ and $|h-h_c|$ dependence separately. For $\vec{q}$ dependence, we found, $\chi^{-1}(T,h_c,\vec{q},0)-\chi^{-1}(T,h_c,0,0)$ can be well fitted by $c_q |\mathbf{q}|^2$ as showed in Fig.~\ref{fig:chifit}(a). In the Fig.~\ref{fig:chiqpower}, we also tried to fit a general form $c_q |\mathbf{q}|^{a_q}$, i.e set the power $a_q$ as free fitting parameter, and obtained $a_q=1.98 \pm 0.06$, which means that the anomalous dimension $\eta =2-a_q=0.02 \pm 0.06$, thus gives a very small upper bound of the numerical value of anomalous dimension.

For $\omega$ dependence, the correction is stronger and $\chi^{-1}(T,h_c,0,\omega)-\chi^{-1}(T,h_c,0,0)$ can be well fitted by $c_\omega \omega + c_\omega'\omega^2$, as showed in Fig.~\ref{fig:chifit}(b).

\begin{figure}[htp!]
\centering
\includegraphics[width=\columnwidth]{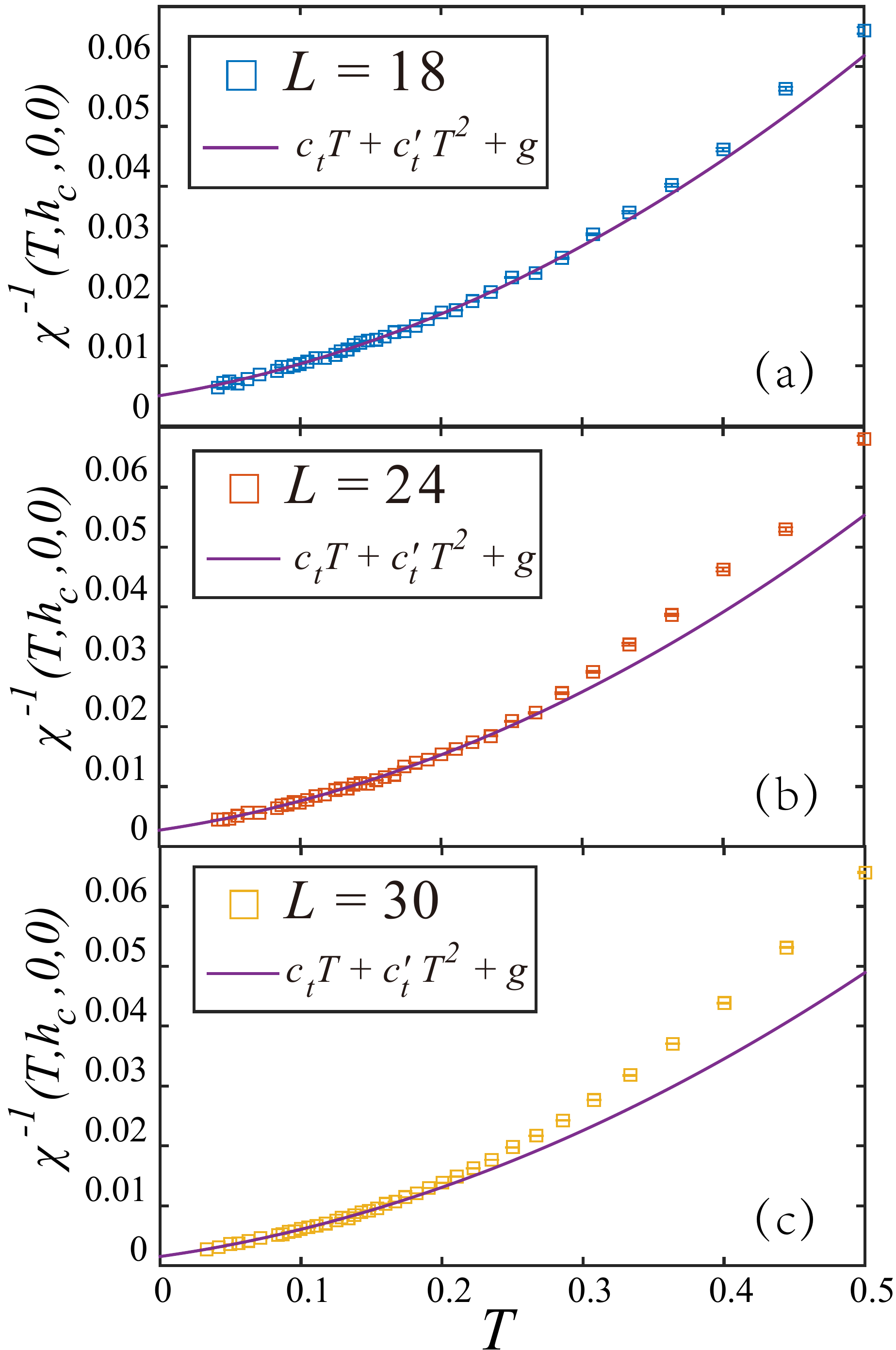}
\caption{$T$ dependence of spin susceptibility $\chi(T,h_c,\mathbf{q},\omega)$ near quantum critical point. (a), (b) and (c) are for system sizes $L=18$, $24$ and $30$, respectively. The obtained fitting parameters are summarized in Table~\ref{table1}.}
\label{fig:chitL}
\end{figure}

For temperature dependence, we found in Fig.~\ref{fig:chifit} (c), when $\vec{q}=0$ and $\omega=0$, the spin susceptibility can be well described by 
%
\begin{align}
\chi(T,h_c,\mathbf{q}=0,\omega_n=0)=\frac{1}{c_{t}T+c_{t}'T^{2}+g},
\label{eq:chiT}
\end{align}
where the background constant $g$ is put to take care of finite size effect. As further elucidated in Fig.~\ref{fig:chitL}, the background constant $g$ reduces as system size $L$ increases. Moreover, the fitting parameter $c_t$ and $c'_t$, summerized in Table~\ref{table1}, can be seen to converge to finite values as $L$ increases. The converged values of $c_t$ and $c'_t$ on the largest $L$ we have simulated is also shown in Fig.~\ref{fig:chifit}(c). 

\begin{table}[ht]
    \caption{The obtained fitting parameters for temperature dependence of spin susceptibilities showed in Fig.~\ref{fig:chitL}.} 
    \centering 
    \def\arraystretch{1.5}
    \begin{tabular}{L{0.8cm}  c C{2cm} c } 
    \hline\hline 
    $L$ & $c_t$ & $c_t'$ & $g$ \\ [0.0ex] 
    \hline 
     18  &  0.038(6)  &  0.15(2)  &  0.0050(5) \\ [0.0ex] 
    \hline                      
     24  &  0.034(5)  &  0.14(1)  &  0.0027(4) \\ [0.0ex] 
    \hline                      
     30  &  0.033(7)  &  0.12(3)  &  0.0015(4) \\ [0.0ex] 
    \hline 
    \end{tabular}
    \label{table1} 
\end{table}

In summary, the spin susceptibility in the presence of itinerant fermions can be described by the following form
\begin{align}
&\chi(T,h,\mathbf{q},\omega_n)= \nonumber\\
& \frac{1}{(c_{t}T+c_{t}' T^{2})+c_{h}|h-h_c|^{\gamma}+c_q |\mathbf{q}|^2 + (c_{\omega}\omega+c'_{\omega}\omega^{2})},
\label{eq:susceptibility}
\end{align}
where the values of the parameters are shown in Fig~\ref{fig:chifit} (a), (b) and (c). This spin susceptibility demonstrates a crossover behavior between the low-energy Hertz-Millis-Moriya universality class and the high-energy (2+1)D XY  universality class. At low energy and temperature ($T \ll c_{t}/c_{t}'$ or $\omega\ll c_{\omega}/c'_{\omega}$), the subleading quadratic terms of $\omega$ and $T$  become negligible and thus the Hertz-Millis-Moriya scaling is observed with the signature linear $T$ and $\omega$ dependence. Here, $c_{t}/c_{t}'$ and $c_{\omega}/c'_{\omega}$ define the crossover scales of temperature and energy, and also serve as the temperature and energy cutoff in the Hertz-Millis-Moriya theory. Above this cutoff, at high $T$ and $\omega$, the quadratic  $T$ and $\omega$ terms dominate, and the  (2+1)D XY exponents are recovered up to a small and unmeasurable anomalous dimension. This result is consistent with the understanding that fermion-boson coupling at this QCP is relevant at infared in RG. As a relevant operator, this coupling becomes stronger (weaker) at low (high) energy, and thus the scaling exponents deviate from (recovers) the bare bosonic thoery.

To verified that these conclusions are not polluted by finite-size effects, we examine the correlation length at the onsite of the cross point $\omega^*=c_{\omega}/c'_{\omega}$ and  $T^*=c_{t}/c_{t}'$. At $\omega^*$ and $T^*$ the correlation length, $\xi\sim \sqrt{c_q \chi(q=0)}$, is about $4$ and $7$ respectively, both significantly smaller than the system size $L=30$, and thus finite-size contributions are well controlled. It is also worthwhile to highlight that the correlation length at $T^*$ is much larger than that at $\omega^*$, implying that the linear $T$ dependences only arises at the close vicinity of the QCP in comparison with the linear $\omega$ behavior. This observation is consistent with the Hertz-Millis-Moriya theory, where the linear $\omega$ behavior comes from the leading order RPA corrections, while the linear $T$ dependence requires high order effect, i.e. running coupling in RG, which only becomes significant near the QCP.

The crossover behavior discussed above has a direct impact on physical quantities. Here, as an example, we consider the temperature dependence of the equal-time spin-spin correlation function at zero wavevector $C(T, \mathbf{q}=0, \tau=0)$. Based on the susceptibility shown in  Eq.~\eqref{eq:susceptibility}, this correlation function shall take the following analytic form
\begin{align}
C(T)=-\frac{T}{\delta}+\frac{\psi\left(\frac{c_{\omega}+\sqrt{c_{\omega}^2-4 c'_{\omega} \delta}}{4\pi c'_{\omega} T}\right)
-\psi\left(\frac{c_{\omega}-\sqrt{c_{\omega}^2-4 c'_{\omega} \delta}}{4\pi c'_{\omega} T}\right)
}{\pi \sqrt{c_{\omega}^2-4 c'_{\omega} \delta}}
\label{eq:equal-time}
\end{align}
where $\psi(x)$ is the digamma function and $\delta=\chi(T,h,\mathbf{q}=0,\omega_n=0)^{-1}=\delta_0+c_{t}T+c_{t}' T^{2}$ is the mass of the boson modes, which measures the distance away from the QCP, with $\delta=0$ at the QCP. We emphasize that here all the control parameters are determined by Eq.~\eqref{eq:susceptibility} and Figs.~\ref{fig:chifit}(a-c), and there is no other adjustable parameter in this theory. As shown in Fig.~\ref{fig:chifit}(d), our numerical data agrees nicely with the analytic result. More importantly,  the asymptotic form of this analytic formula demonstrates clearly a  crossover behavior between the low-$T$ $z=2$ (Hertz-Millis-Moriya) and the high-$T$ $z=1$ ((2+1)D XY) critical scalings. As can be seen in Fig.~\ref{fig:chifit}(d), at high temperature, $C(T)\sim\coth(\sqrt{\delta}/2 \sqrt{c'_{\omega}} T)/2\sqrt{c'_{\omega} \delta}$, consistent with a QCP with $z=1$ and the linear $\omega$ term in spin susceptibility plays no role here. At low $T$, however, $C(T)\sim \frac{1}{\pi \sqrt{c_\omega^2-4 c'_\omega \delta}}\log(\frac{c_\omega+\sqrt{c_\omega^2-4 c'_\omega \delta}}{c_\omega-\sqrt{c_\omega^2-4 c'_\omega \delta}})+c_\omega \pi T^2/3\delta^2$. In the close vicinity of the QCP ($\delta \to 0$ and $T\to 0$), this low $T$ asymptotic form becomes 
$C(T)\sim \frac{\log (\Lambda/\delta)}{\pi \sqrt{c_\omega}}$ with ultraviolet cutoff $\Lambda= c_\omega^2/c'_\omega$. This logarithmic behavior is the key signature of the $z=2$ QCP.

\section{Discussions}
In this paper, we examine the QCP with itinerant fermions. For order parameters with a finite wavevector, we find that the low energy scaling behavior agrees with the Hertz-Millis-Moriya theory, and the fermion contributions becomes irrelevant at high energy. We have also provide an upper bound of the possible anomalous  dimension predicted for this QCP beyond the Hertz-Millis-Moriya setting.

Although the linear $T$ behavior that we observed is consistent with the RG theory of Millis~\cite{Millis1993}, it is worthwhile to emphasize that in the Hertz-Millis-Moriya theory, this linear $T$ contribution is predicted for extremely low temperature limit, $\ln[\ln(1/T)] \gg 1$. As pointed out in Ref.~\cite{Sachdev2006}, this assumption is essentially impossible to satisfy in practice and our simulations cannot access such an extremely low-$T$ limit either. Instead, our results indicate that the linear $T$ behavior near such a QCP survives to much higher temperature beyond $\ln[\ln(1/T)] \gg 1$. Such a stable linear $T$ behavior (at higher $T$) is consistent with the model analyzed in Ref.~\cite{Sachdev2006},which also contains a linear $T$ correction but with a different logarithmic correction. In principle, one can differentiate the linear $T$ correction in Ref.~\cite{Sachdev2006} with that in the Hertz-Millis-Moriya theory by examining the sub-leading logarithmic corrections. However, such a delicate analysis is highly challenging for numerical studies and we leave it for future investigations.

\section*{Acknowledgments}
We acknowledge valuable discussions with Sung-Sik Lee and Emanuel Gull.
ZHL, XYX and ZYM acknowledge fundings from the Ministry of Science and Technology of China through National Key Research and Development Program under Grant No. 2016YFA0300502 and the key research program of the Chi-
nese Academy of Sciences under Grant No. XDPB0803 and
from the National Science Foundation of China under Grant
Nos. 11421092, 11574359 and 11674370 as well as the National Thousand-Young Talents Program of China. We thank the Center for Quantum Simulation Sciences in the Institute of Physics, Chinese Academy of Sciences and the Tianhe-1A platform at the National Supercomputer Center in Tianjin for their technical support and generous allocation of CPU time. XYX also acknowledges the support of HKRGC through grant C6026-16W. K.S. acknowledges support from the National Science Foundation under Grant No. PHY1402971 and the Alfred P. Sloan Foundation. Y.Q. thanks the DOE Office of Basic Energy Sciences, Division of Materials Sciences and Engineering under Award de-sc0010526. Y.Q. is also supported by the Ministry of Science and Technology of China under Grant No. 2015CB921700.

\appendix

%
%

\section{Self-learning quantum Monte Carlo method}
\label{sec:SLDQMCIntroduction}

In this work, we performed the recently developed self-learning determinantal quantum Monte Carlo (SLDQMC) simulations~\cite{Xu2016self} to speed up our simulations. Before introduce details of SLDQMC, we first briefly explain the DQMC framework for our model.
  In DQMC, the partition function is expressed as
\textbf{
\begin{eqnarray}
Z & = & \text{Tr}\left[e^{-\beta\hat{H}}\right]\nonumber\\
 & = & \sum_{s^{z}_{1}\cdots s^{z}_{N}=\pm1}\text{Tr}_{F}\left\langle s^{z}_{1}\cdots s^{z}_{N}\left|\left(e^{-\Delta\tau\hat{H}}\right)^{M}\right|s^{z}_{1}\cdots s^{z}_{N}\right\rangle
\end{eqnarray}
}
where $\vec{S}=\left(s^{z}_{1}\cdots s^{z}_{N}\right)$ denoting the Ising spins, then
\begin{align}
Z =&\sum_{\vec{S}_{1}\cdots\vec{S}_{M}}\text{Tr}_{F}\langle\vec{S_{1}}|e^{-\Delta\tau\hat{H}}|\vec{S}_{M}\rangle\times\nonumber\\
&\langle\vec{S_{M}}|e^{-\Delta\tau\hat{H}}|\vec{S}_{M-1}\rangle\cdots\langle\vec{S_{2}}|e^{-\Delta\tau\hat{H}}|\vec{S}_{1}\rangle,
\end{align}
now we can trace out the fermion degrees of freedom, and obtain the configurational weight,
\begin{equation}
\omega_{\mathcal{C}}=\omega_{\mathcal{C}}^{TI}\omega_{\mathcal{C}}^{F}
\end{equation}
with the Ising part
\begin{equation}
\omega_{\mathcal{C}}^{TI} =  \left(\prod_{\tau}\prod_{\langle i,j\rangle}e^{\Delta\tau Js^{z}_{i,\tau}s^{z}_{j,\tau}}\right)\left(\prod_{i}\prod_{\langle\tau,\tau'\rangle}\Lambda e^{\gamma s^{z}_{\tau,i}s^{z}_{\tau',i}}\right)
\end{equation}
where $\Lambda^{2}=\sinh(\Delta\tau h)\cosh(\Delta\tau h)$, $\gamma=-\frac{1}{2}\ln\left(\tanh(\Delta\tau h)\right)$.
For the fermion part, we have
\begin{equation}
\omega_{\mathcal{C}}^{F} =  \det\left(\mathbf{1}+\mathbf{B}_{M}\cdots\mathbf{B}_{1}\right)
\end{equation}
As an anti-unitary symmetry $i\tau_yK$ (where $\tau_y$ is a Pauli matrix in the orbital basis and $K$ is the complex conjugation operator) 
 make the Hamiltonian invariant, the
fermion part ratio can be further rewritten as
\begin{equation}
\omega_{\mathcal{C}}^{F}=\left|\prod_{\sigma}\det\left(\mathbf{1}+\mathbf{B}_{M}^{1\sigma}\cdots\mathbf{B}_{1}^{1\sigma}\right)\right|^{2}
\end{equation}
where
\begin{equation}
\mathbf{B}_{\tau}^{\lambda\sigma}=\exp\left(-\Delta\tau\mathbf{K}^{\lambda\sigma}+\Delta\tau\xi\text{Diag}(s_{1}^{z},\cdots,s_{N}^{z})\right)
\end{equation}
with $\mathbf{K}^{\lambda\sigma}$ the hopping matrix for orbital $\lambda$ and spin $\sigma$. It turns out both the fermion weight and the Ising weight are always positive, thus there is no sign problem. To systematically improve the simulation, especially close to (quantum) critical point, we have implemented both local update in DQMC and space-time global update~\cite{Xu2016a}. In the global update, we use Wolff algorithm~\cite{Wolff1989} and geometric cluster algorithm~\cite{Heringa1998} to propose space-time clusters of the Ising spins and then calculate the fermion weight to respect the detail balance as the acceptance rate of the update. 

Since we are interested in the properties of the system in the quantum critical region, strong autocorrelations in the DQMC simulations are expected. To overcome this problem, we furthermore performed the self-learning determinantal DQMC, dubbed SLDQMC~\cite{Xu2016self}, in which, to reduce the autocorrelation and speedup the simulation, we use cumulative update with bosonic effective model self-learnt from the feedback of fermions via Ising spin configurations generated with the above mentioned update scheme for the original model.
 The SLDQMC algorithm is made up of following steps.

(i) Use the local update plus Wolff and geometric cluster updates with DQMC to generate enough configurations according to the original Hamiltonian. 

(ii) Obtain an effective model by self-learning process~\cite{Xu2016self,liu2016self,liu2016fermion}. The effective model can be very general,
\begin{equation}
H^{\text{eff}} =  E_0 +  \sum_{ (i\tau);(j,\tau')}J_{i,\tau;j\tau'} s_{i,\tau} s_{j,\tau'}+\cdots
\label{eq:effectiveHam}
\end{equation}
where $J_{i,\tau;j\tau'}$-s parameterize the two-body interaction between any bosonic field in space-time. More-body interactions, denoted as $\cdots$, can also be included, although not in this Letter.

(iii) Perform multiple local updates with $H^{\text{eff}}$ (as in general the $H^{\text{eff}}$ will contain non-local terms which make the cluster update difficult to implement). Given a configuration $\mathcal{C}$ and $\mathcal{C'}$ is the configuration after cumulative local update, the acceptance ratio of the cumulative update can be derived from the detail balance as
\begin{equation}
 A(\mathcal{C} \rightarrow \mathcal{C}') = \min \left\{1, \  \frac {\exp\left(-\beta H[\mathcal{C}'] \right) } { \exp\left(-\beta H[\mathcal{C}]\right)}  \frac {\exp\left(-\beta H^{\text{eff}} [\mathcal{C}] \right)} {\exp\left(-\beta H^{\text{eff}} [\mathcal{C}'] \right)}  \right\}.
\label{eq:acceptanceratio}
\end{equation}
this entire process is denoted as a {\it cumulative update}. Since the effective model is very close to the low-energy description of the original Hamiltonian, the cancellation in the exponential factors in Eq.~\ref{eq:acceptanceratio} can easily give rise to acceptance ratio very close to 1.

(iv) following the detailed balance decision, we decide to accept or reject the final configuration via evaluating the fermionic determinant.

Different from the local update in DQMC, the local move of $H^{\text{eff}}$ in step (iii) is very fast, as there are no matrix operations associated with DQMC involved. Furthermore, to generate statistically independent configurations at (quantum) critical point, the number of sweeps of local update should be comparable with correlation time $\tau_L$. With these local updates of effective model, the configuration has been changed substantially, and we take the final configuration as a proposal for a global update for the original model. With these efferots, we now can simulate systems with $L=30$ and $\beta=40$ comfortably. To our knowledge, it is the new record in the finite temperature DQMC simulation.

\section{Correlation ratio close to QCP}
\label{sec:correlationratio}
\begin{figure}[htp!]
\centering
\includegraphics[width=\columnwidth]{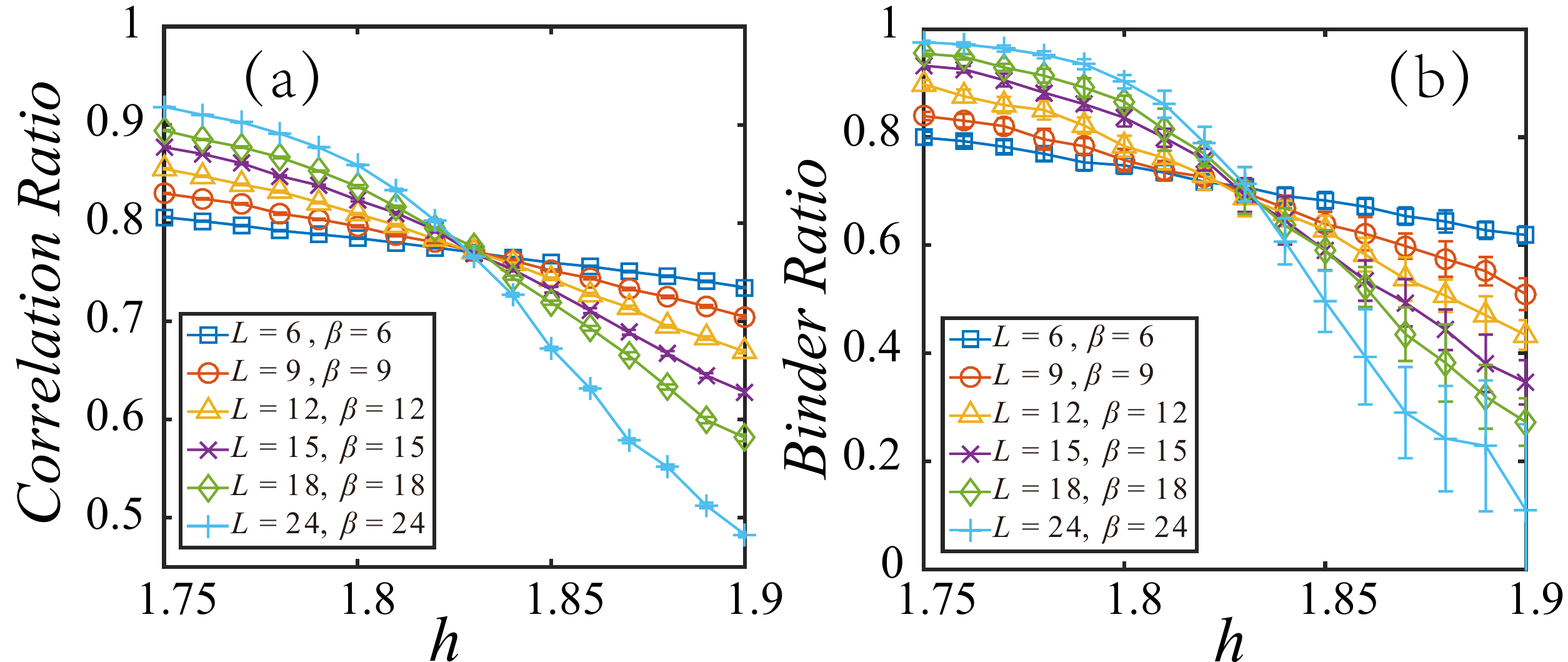}
\caption{(a)Correaltion ratio vs transverse field with $\beta=L$ scaling relation and (b)Binder Ratio vs transverse field with $\beta=L$ scaling relation. The crossing point of the correlation ratio $R_c$ and the Binder Ratio $R_b$ converge to $h_c=1.83(1)$ which is the quantum critical point of our model.}
\label{fig:Rc-ratio}
\end{figure}

To determine the precise position of the QCP, we implement the measurement of correlation ratio and binder ratio of the Ising spin-spin correlation function~\cite{Pujari2016}. Correlation ratio $R_c$ is defined from spin susceptibility at zero frequency. 
\begin{equation}
R_c=1-\frac{\chi(T,h,\mathbf{k}=\mathbf{Q}+\mathbf{b}/L,\omega=0)}{\chi(T,h,\mathbf{k}=\mathbf{Q},\omega=0)},
\end{equation}
where $\mathbf{b}$ is the shortest reciprocal lattice vector. And Binder Ratio is defined from Ising spin order parameter $m$.
\begin{equation}
R_b=2-\frac{<m^4>}{<m^2>^2},
\end{equation}

As shown in Fig.~\ref{fig:Rc-ratio}, the crossing points of the $R_c$ and $R_b$ determine the position of the QCP. Here we have purposely chosen the scaling relation between system size and inverse temperature as $\beta=L$. The crossing point of $R(L)$ points to the critical point at $h_c=1.83(1)$. 

\section{Quantum critical scaling analysis of bosonic susceptibility in the absence of fermions}
\label{sec:bosonicsuscepU0}
\begin{figure}[htp]
\centering
\includegraphics[width=\columnwidth]{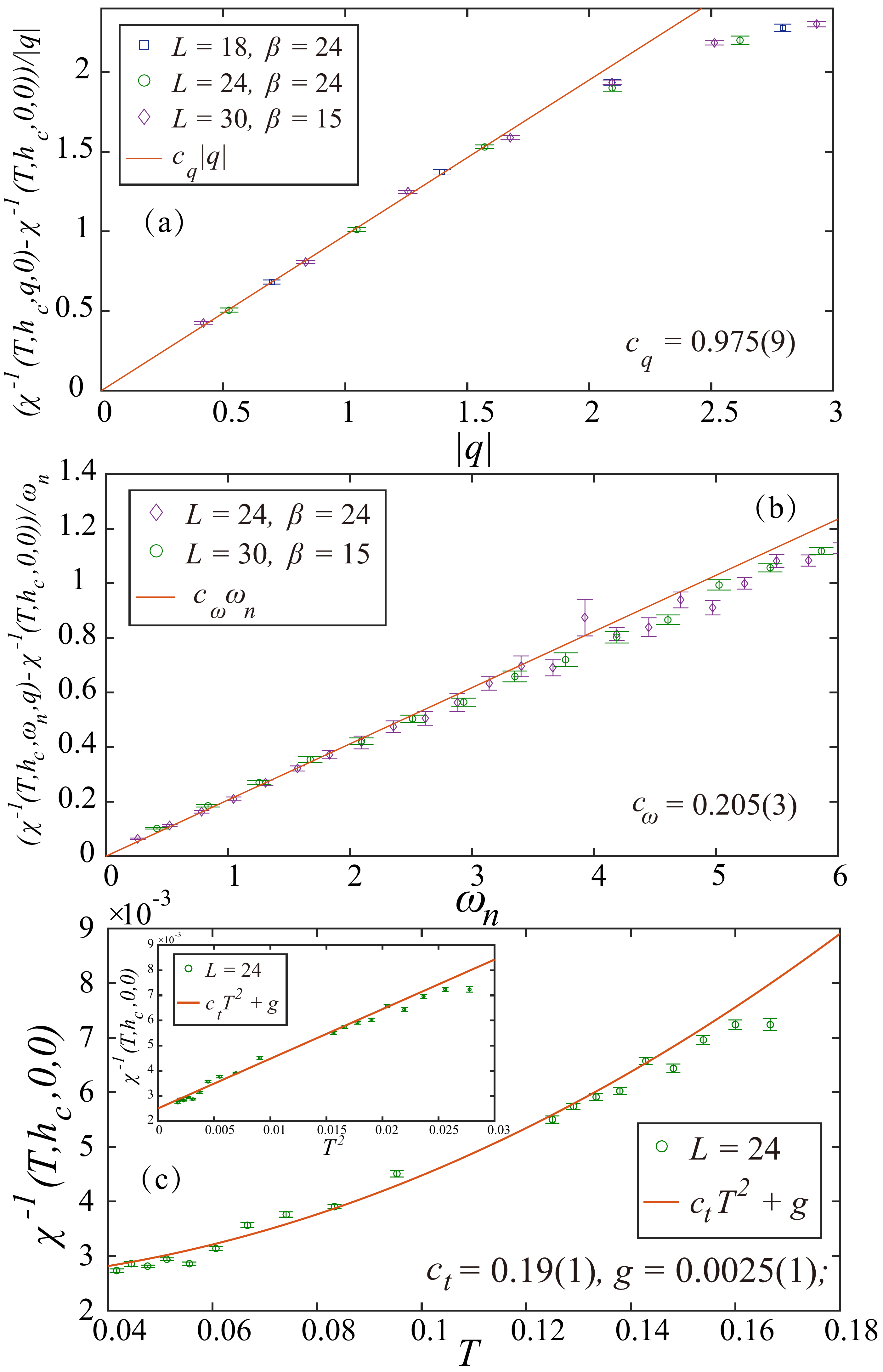}
\caption{(a) $\mathbf{q}$ dependance of the 2+1D XY dynamic bosonic susceptibility $\frac{\chi^{-1}(\mathbf{q})-\chi^{-1}(0)}{|\mathbf{q}|}$. (b) $\omega$ dependance of the 2+1D XY dynamic bosonic susceptibility $\frac{\chi^{-1}(\omega)-\chi^{-1}(0)}{\omega}$. (c) $T$ dependence of the 2+1D XY dynamic bosonic susceptibility $\chi^{-1}(T,h_c,|\mathbf{q}|=0,\omega=0)$. The fits are performed according to Eq.~\ref{eq:XYsusceptibility}.}
\label{fig:chi-analysis-U0}
\end{figure}

In the absence of fermions, the QCP is in the 2+1D XY universality class. In the main text, we discussed that the Ising spin susceptibility takes the following form (in the approximation $\eta=0$)

\begin{align}
\chi(T,h,\mathbf{q},\omega_n)= \frac{1}{c_{t}T^2+c_{h}|h-h_c|^{\gamma}+c_q |\mathbf{q}|^2 + c_{\omega}\omega^{2}},
\label{eq:XYsusceptibility}
\end{align}
where the main difference with the case with fermion is the only quadartic form of $\omega$ and $T$ dependence in $\chi^{-1}$. Here, we plot the bare 2+1D XY model susceptibility in QCP $h_c=1.63(1)$ and the values of coefficients are determined from scaling analyses, as shown in Fig.~\ref{fig:chi-analysis-U0}.

In Fig~\ref{fig:chi-analysis-U0} (a) and (b), we plot $(\chi^{-1}(\mathbf{q})-\chi^{-1}(0))/|\mathbf{q}|$ and $(\chi^{-1}(\omega)-\chi^{-1}(0))/{\omega}$ which show obivious linear behavior. Although the $\chi^{-1}$ vs $\mathbf{q}$ relation is very similar with fermion coupling case in the main text, the $\chi^{-1}$ vs $\omega$ relation show no intersection at zero frequency, indicating only quadartic frequency dependence in bare XY model. In Fig~\ref{fig:chi-analysis-U0} (c), the quadartic temperature relation fits the data well. The inset of Fig~\ref{fig:chi-analysis-U0} (c)  plot $\chi^{-1}$ as function of $T^2$ and the data show good linear behavior which is consistent with $T^2$ dependance for the 2+1D XY universality and different from $T+T^2$ dependance of the coupled case in the main text.

\section{Superfluid Density}
\label{sec:superfluiddensity}

\begin{figure}[htp]
\centering
\includegraphics[width=\columnwidth]{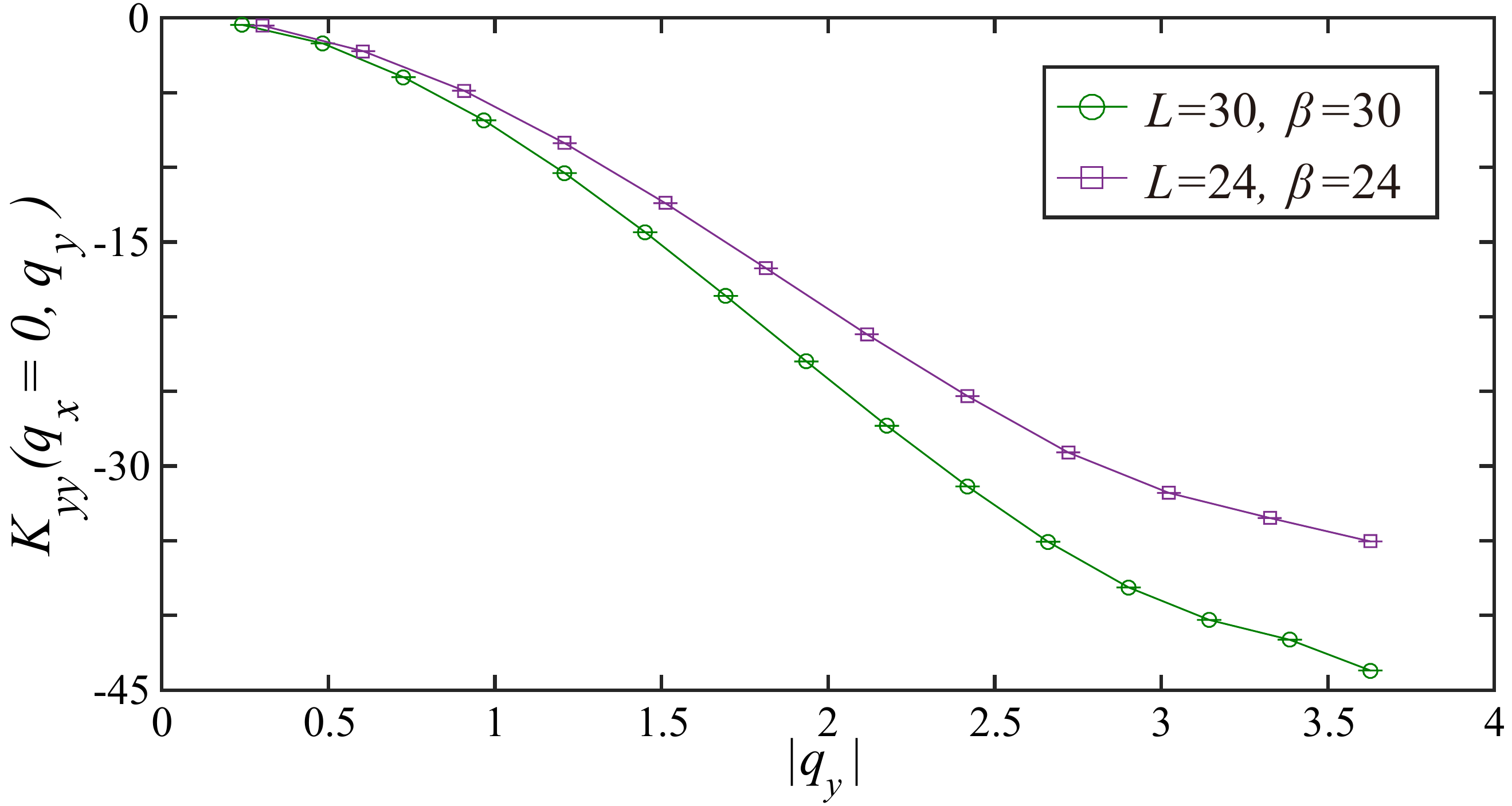}
\caption{$K_{yy}(q_x=0,q_y)$ for different temperature at $h=h_c$ for a $L=24$ and $L=30$ system. When $q_y$ approaches zero, the superfluid density approaches a negative number, indicating a paramagnetic response, and hence there is no strong superconductivity fluctuations.}
\label{fig:superfluid-density}
\end{figure}

To test whether there are possible superconductivity instabilities close to the QCP, we measured the superfluid density, following literature~\cite{Xu2016}, it is given by
\begin{equation}
\rho_{s}=\lim_{q_{y}\rightarrow0}\lim_{L\rightarrow\infty}K_{yy}(q_{x}=0,q_{y})
\end{equation}
with
\begin{equation}
K_{yy}(\vec{q})=\frac{1}{4}\left[\Lambda_{yy}(q_{x}\rightarrow0,q_{y}=0)-\Lambda_{yy}(\vec{q})\right]
\end{equation}
and the current-current correlation function,
\begin{equation}
\Lambda_{yy}(\vec{q})=\sum_{i}\int_{0}^{\beta}d\tau e^{-i\vec{q}\cdot\vec{r}_{i}}\langle j_{y}(\vec{r}_{i},\tau)j_{y}(0,0)\rangle,
\end{equation}
where $\delta$ represent the nearest neighbor hopping direction and $\delta_y$ is its $y$ component, where
\begin{equation}
j_{y}(\vec{r}_{i}) = it\sum_{\lambda\sigma\delta}\delta_{y}\hat{c}_{i\lambda\sigma}^{\dagger}\hat{c}_{i+\delta,\lambda\sigma}.
\end{equation}

As shown in Fig.~\ref{fig:superfluid-density}, at $h\approx h_c$ and as one goes down in temperature, our measurement of $K_{yy}$ shows that close to QCP, the superfluid density is always negative even when the temperature is as low as to $\beta=30$ and system size as large as $L=30$. Hence down to $\beta=30$, we do not find signature of superfluid density and correspondingly no emergent superconductivity.

\bibliography{triangle-qcp}

\end{document}